# High Performance Optical Filters Using Three Waveguide Coupled Sagnac Loop Reflectors

H.Arianfard, J.Wu, *Member, IEEE*, S.Juodkazis, *Fellow, OSA*, D.J. Moss, *Fellow, IEEE, Fellow, OSA*

*Abstract*—We theoretically investigate advanced multi-functional integrated photonic filters formed by three waveguide coupled Sagnac loop reflectors (3WC-SLRs). By tailoring the coherent mode interference, the spectral response of the 3WC-SLR resonators is engineered to achieve diverse filtering functions with high performance. These include optical analogues of Fano resonances that yield ultrahigh spectral extinction ratios (ERs) and slope rates, resonance mode splitting with high ERs and low free spectral ranges, and classical Butterworth, Bessel, Chebyshev, and elliptic filters. A detailed analysis of the impact of the structural parameters and fabrication tolerances is provided to facilitate device design and optimization. The requirements for practical applications are also considered. These results theoretically verify the effectiveness of using 3WC-SLR resonators as multi-functional integrated photonic filters for flexible spectral engineering in diverse applications.

*Keywords*—Integrated optics, resonators, Fano resonance, mode splitting, classical filters.

## I. INTRODUCTION

IN concert with the advance of complementary metal-oxide-semiconductor (CMOS) fabrication technologies, particularly as applied to photonic integrated circuits, integrated photonic resonators have become key building blocks [1, 2]. With narrowband wavelength selectivity, strong resonance field enhancement, and versatile filter shapes, these micro/nano-scale devices have been widely used for diverse applications including filters, lasers, modulators, buffers, switches, sensors, and signal processors [2-10]. and have also been applied to phase-sensitive all-pass filters [11-15]. Integrated ring resonators, particularly in CMOS compatible platforms, have also formed the basis for Kerr microcombs, [16-31] that have experienced wide applications to microwave photonics, [32-55] data communications [56-58] and neural networks [59, 60], and quantum optics [61- 69]. Photonic crystal cavities form a key class of integrated resonator that have been widely investigated [70-76]. For all of these devices, CMOS compatible platforms have proven to be indispensable [77 - 88] for their low loss, reproducibility and manufacturability, particularly for nonlinear optical applications where they have out-performed silicon [89-103] and even chalcogenide glass [104-117] and even other materials [118-120] because of their low nonlinear absorption [121-123]. This has been a motivation for integrating novel 2D materials onto CMOS platforms, to increase their nonlinear performance [124 - 130].

Fano resonances are a fundamental physical phenomenon featuring an asymmetric resonant lineshape profile induced by interference between a discrete localized state and a continuum state [131-134]. It was first discovered in the absorption spectra of noble gases and later on has been extended to a much wider scope [135, 136]. In recent years, various types of photonic resonators have been designed to realize optical analogues of Fano resonances, which have attracted great interest and found wide applications in optical switching, sensing, light focusing beyond the diffraction limit, topological optics, data storage, and many others [[131, 133, 137-139].

Resonance mode splitting is a fundamental phenomenon in photonic resonators that occurs when two or more mutually coupled modes co-exist in the same resonant cavity [140, 141]. It can achieve a reduced free spectral range (FSR) and an increased quality (Q) factor while maintaining a small physical cavity length, thus yielding a compact device footprint, low power consumption, and versatile filter shapes for dense-wavelength division-multiplexing (DWDM) and microwave photonics applications [142, 143]. Recently, many applications based on mode-split resonators have been demonstrated, such as optical buffering [144], dispersion compensation [145,146], signal multicasting [147, 148], differential equation solving [149], microwave signal generation [150], and sensing [151].

Butterworth, Bessel, Chebyshev, and elliptic filters are classical filters that model and govern signal filtering and processing in communications and computing systems [152-154]. These fundamental filters are broadly used in applications such as noise reduction, spectral analysis, and signal synthesis [155-158]. Photonic resonators offer a powerful solution to realize these filters in optical domain [159-161]. With much broader operation bandwidth than their electronic counterparts, these optical filters play an important role in high-speed optical communications and information processing [162].

To realize Fano-resonances, resonance mode splitting, and classical filters based on integrated photonic resonators could offer competitive advantages including compact footprint, high scalability, high stability, and mass producibility for practical applications. Recently, we investigated multi-functional integrated photonic filters based on cascaded Sagnac loop reflectors (SLRs) [163] and two waveguide coupled Sagnac loop reflectors (2WC-SLRs) [164] formed by

The authors are with the Optical Sciences Centre, Swinburne University of Technology, Hawthorn, VIC 3122, Australia (e-mail: dmoss@swin.edu.au).

self-coupled silicon-on-insulator (SOI) nanowires. Here, we theoretically investigate more complex device structures including three waveguide coupled Sagnac loop reflectors (3WC-SLRs) that yield greatly enhanced performance together with more versatile filtering functions. We tailor coherent mode interference in the 3WC-SLR resonators to achieve versatile filter shapes with high performance, including optical analogues of Fano resonances with ultrahigh extinction ratios (ERs) and slope rates (SRs), resonance mode splitting with high ERs and low FSRs, and classical Butterworth, Bessel, Chebyshev, and elliptic filters. A detailed analysis of the impact of the structural parameters and fabrication tolerances is provided. The requirements for practical applications are also considered in our design and achieved for the proposed devices. These results highlight the strong potential of 3WC-SLR resonators for flexible spectral engineering in diverse applications such as optical filtering, switching, and sensing.

## II. DEVICE CONFIGURATION

Figure 1 illustrates schematic configurations of the 3WC-

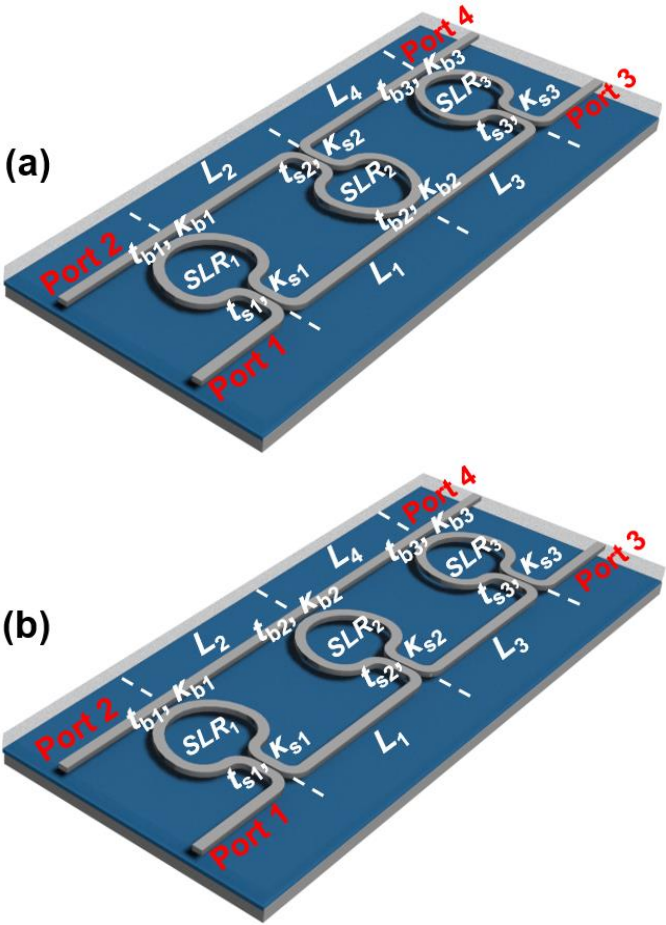

Fig. 1. Schematic configuration of (a) zig-zag and (b) parallel 3WC-SLR resonators consisting of three SLRs ($SLR_1$, $SLR_2$, and $SLR_3$), respectively. The definitions of $t_{si}$ ($i$ = 1, 2, 3), $t_{bi}$ ($i$ = 1, 2, 3), $L_{SLRi}$ ($i$ = 1, 2, 3), and $L_i$ ($i$ = 1, 2, 3, 4) are given in Table I.

SLR resonators. Two types of 3WC-SLR resonators are investigated. The first one in Fig. 1(a) consists of three inversely coupled SLRs and is termed a zig-zag 3WC-SLR resonator, while the second one in Fig. 1(b) consists of three parallel Sagnac loop reflectors (SLRs) coupled to a top bus waveguide, and is termed a parallel 3WC-SLR resonator. In both 3WC-SLR resonators, the bus waveguides between the SLRs introduce additional feedback paths for coherent optical mode interference, yielding greatly improved flexibility for engineering the spectral response. To study the 3WC-SLR resonators based on the scattering matrix method [163-165], the waveguide and coupler parameters are defined in Table I. To simplify the comparison, we assume that the three SLRs are identical in each 3WC-SLR resonator, i.e., $L_{SLR1}$ = $L_{SLR2}$ = $L_{SLR3}$ = $L_{SLR}$, $t_{s1}$ = $t_{s2}$ = $t_{s3}$ = $t_s$, and $t_{b1}$ = $t_{b2}$ = $t_{b3}$ = $t_b$.

The zig-zag 3WC-SLR resonator is equivalent to two cascaded Mach–Zehnder interferometers (MZIs, which is a finite-impulse-response (FIR) filter) and a Fabry-Perot cavity (which is an infinite-impulse-response (IIR) filter) when $t_s$ = 1 and $t_b$ = 1, respectively. On the other hand, the parallel 3WC-SLR resonator is equivalent to two cascaded MZIs and three cascaded SLRs (which is an IIR filter) when $t_s$ = 1 and $t_b$ = 1, respectively. When $t_s \neq 1$ and $t_b \neq 1$, both types of filters can be considered to be hybrid, consisting of both FIR and IIR filter elements that allows more versatile coherent mode interference induced by mutual interaction. As compared with the 2WC-SLR resonators [164], the 3WC-SLR resonators have an extra SLR and additional feedback paths that introduce more complex coherent mode interference, which can lead to enhanced filter performance and versatility. The freedom in designing the reflectivity of the SLRs (i.e., $t_s$), the coupling strength between the SLRs and connecting bus waveguides (i.e., $t_b$), and the waveguide lengths (i.e., $L_{SLR}$ and $L_i$) enables flexible spectral engineering based on the 3WC-SLR resonators, which can lead to diverse filtering functions.

In the following sections, mode interference in the 3WC-SLR resonators is tailored to achieve high-performance filtering functions, including optical analogues of Fano resonances (Section III), resonance mode splitting (Section IV), and classical Butterworth, Chebyshev, Bessel, and elliptic filters (Section V). In our design, we use values obtained from our previously fabricated SOI devices [163, 166] for the waveguide group index ($n_g$ = 4.3350, transverse electric (TE) mode) and the propagation loss ($\alpha$ = 55 m$^{-1}$, i.e., 2.4 dB/cm). The devices are designed based on but not limited to the SOI integrated platform.

TABLE I
DEFINITIONS OF STRUCTURAL PARAMETERS OF THE 3WC-SLR RESONATORS

| Waveguides | Length | Transmission factor [a] | Phase shift [b] |
|---|---|---|---|
| Bus waveguides between SLRs ($i$ = 1, 2, 3, 4) | $L_i$ | $a_i$ | $\varphi_i$ |
| Sagnac loop in $SLR_i$ ($i$ = 1, 2, 3) | $L_{SLRi}$ | $a_{si}$ | $\varphi_{si}$ |
| **Directional couplers** | Field transmission coefficient [c] | Field cross-coupling coefficient [c] | |
| Coupler in $SLR_i$ ($i$ = 1, 2, 3) | $t_{si}$ | $k_{si}$ | |
| Coupler between $SLR_i$ and bus waveguide ($i$ = 1, 2, 3) | $t_{bi}$ | $k_{bi}$ | |

[a] $a_i$ = exp (-$\alpha L_i$ / 2), $a_{si}$ = exp (-$\alpha L_{SLRi}$ / 2), $\alpha$ is the power propagation loss factor.
[b] $\varphi_i$ = $2\pi n_g L_i / \lambda$, $\varphi_{si}$ = $2\pi n_g L_{SLRi} / \lambda$, $n_g$ is the group index and $\lambda$ is the wavelength.
[c] $t_{si}^2 + \kappa_{si}^2 = 1$ and $t_{bi}^2 + \kappa_{bi}^2 = 1$ for lossless coupling are assumed for all the directional couplers.

## III. ULTRA-SHARP FANO RESONANCES

In this section, we tailor the spectral response of the zig-zag 3WC-SLR resonator to realize optical analogues of Fano resonances with high ERs and SRs. The power transmission spectrum from Port 2 to Port 4 of the zig-zag 3WC-SLR resonator is depicted in Fig. 2(a). The device structural parameters are $L_{SLR} = L_{1,2,3,4} = 115$ µm, $t_s = 0.743$, and $t_b = 0.994$. One can see that there are periodical Fano resonances with identical asymmetric resonant line-shapes in each period. The FSR is about 200 GHz, which equals the sum of the two channel spacings (CS1 and CS2). The two CSs are very close to each other (CS1 = 101.71 GHz and CS2 = 98.88 GHz), reflecting the high SR of the Fano resonances.

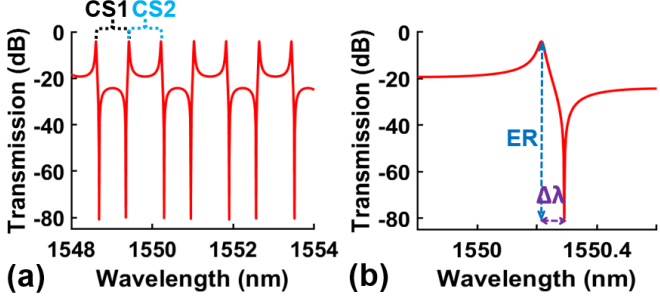

Fig. 2. (a) Power transmission spectrum of the zig-zag 3WC-SLR resonator from Port 2 to Port 4 when $L_{SLR} = L_{1,2,3,4} = 115$ µm, $t_s = 0.743$, and $t_b = 0.994$. (b) Zoom-in view of (a) in the wavelength range of 1549.8 nm –1550.65 nm. CS: channel spacing. ER: extinction ratio. Δλ: wavelength difference between the resonance peak and notch.

Figure 2(b) shows a zoom-in view of Fig. 2(a) in the wavelength range of 1549.8 nm – 1550.65 nm, which shows a Fano resonance with an ultra-high ER of 76.32 dB and an ultra-high SR of 997.66 dB/nm. The ER is defined as the difference between the maximum and the minimum transmission, and the SR is defined as the ratio of the ER to the wavelength difference between the resonance peak and notch (i.e., Δλ in Fig. 2(b)). The high ER and SR reflect the high performance of the Fano resonances resulting from strong coherent optical mode interference in the compact resonator with only three SLRs. Table II compares the performance of the Fano resonances generated by the parallel 2WC-SLR resonator [164] and the zig-zag 3WC-SLR resonator. As compared with the parallel 2WC-SLR resonator, the zig-zag 3WC-SLR resonator can generate Fano resonances with increased ER and SR as well as decreased insertion loss (IL), all of which are highly desirable in practical applications. Moreover, the periodical filter shape of the zig-zag 3WC-SLR resonator is also useful for applications in wavelength division multiplexed (WDM) systems.

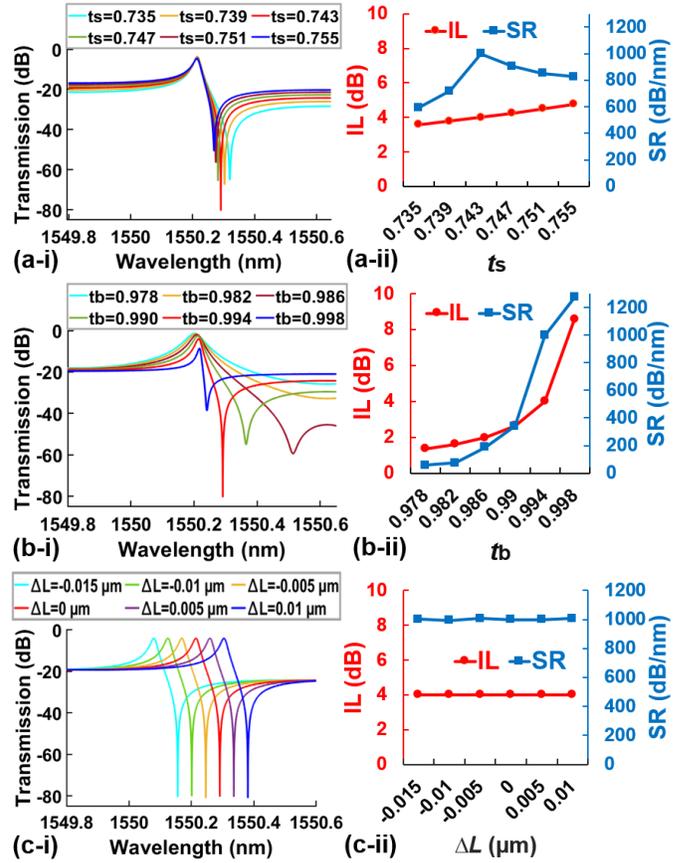

Fig. 3. (a-i) Power transmission spectra and (a-ii) the corresponding SR and IL for various $t_s$ when $t_b$ = 0.994 and $L_{SLR} = L_{1, 2, 3, 4}$ = 115 µm, respectively. (b-i) Power transmission spectra and (b-ii) the corresponding IL and SR for various $t_b$ when $t_s$ = 0.743 and $L_{SLR} = L$ = 115 µm, respectively. (c-i) Power transmission spectra and (c-ii) the corresponding IL and SR for different $L = L_{1,2,3,4}$ when $t_s$ = 0.743, $t_b$ = 0.994, and $L_{SLR}$ = 115 µm, respectively.

TABLE II
PERFORMANCE COMPARISON OF PARALLEL 2WC-SLR AND ZIG-ZAG 3WC-SLR RESONATORS IN ACHIEVING FANO RESONANCES

| Device | IL (dB) | ER (dB) | SR (dB/nm) | FSR (GHz) | Ref. |
|---|---|---|---|---|---|
| Parallel 2WC-SLR resonator [a] | 6.15 | 13.76 | 416.96 | 601.8 | [164] |
| Zig-zag 3WC-SLR resonator | 4 | 76.32 | 997.66 | 200.59 | This work |

[a] The structural parameters are $L_{SLRi} = L_i = 115$ µm ($i$ = 1, 2), $t_s$ = 0.743, and $t_b$ = 0.94.

In Figs. 3(a) − (c), we further investigate the impact of the device structural parameters including $t_s$, $t_b$, and $L = L_{1,2,3,4}$ on the performance of the Fano resonances generated by the zig-zag 3WC-SLR resonator. In each figure, we changed only one structural parameter, keeping the others the same as those in Fig. 2. Figure 3(a-i) compares the power transmission spectra for various $t_s$. The corresponding IL and SR are depicted in Fig. 3(a-ii). The IL increases with $t_s$, while the SR first increases and then decreases with $t_s$, achieving a maximum value of 997.66 dB/nm at $t_s$ = 0.743. The non-monotonic relationship between the SR and $t_s$ is a combined result of both a decrease in Δλ and a non-monotonic variation in ER. The latter mainly arises from the difference between the internal (transmission) and external (coupling) cavity loss, which is similar to that for different coupling regimes in microring resonators (MRRs) [167]. Figure 3(b-i) compares the power transmission spectra for various $t_b$. The IL and SR functions of $t_b$ are depicted in Fig. 3(b-ii). Both the IL and SR increase with $t_b$, reflecting a trade-off between them. Note that although the ER for $t_b$ = 0.994 is higher than for $t_b$ = 0.998, the SR for $t_b$ = 0.994 is higher than that for $t_b$ = 0.998 due to a more significantly decreased Δλ. In Figs. 3(c-i) and (c-ii), we compare the corresponding results for various Δ$L$, which is the length variation of the connecting bus waveguides. To simplify the comparison, we assume the same Δ$L$ for each connecting bus waveguides $L_{1,2,3,4}$ and keep $L_{SLR}$ constant. As Δ$L$ increases, the IL and SR remain unchanged while the resonance redshifts. This highlights the high fabrication tolerance and also indicates that the resonance wavelength can be tuned by introducing thermo-optic micro-heaters [149, 168, 169] or carrier-injection electrodes [142] along the connecting bus waveguides to tune the phase shift.

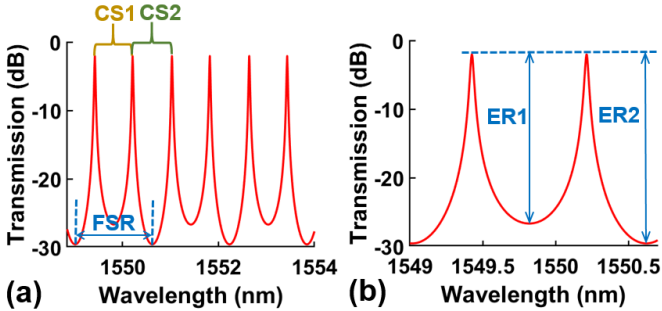

Fig. 4. (a) Power transmission spectrum of the zig-zag 3WC-SLR resonator from Port 2 to Port 4 when $L_{SLR} = L_{1,2,3,4}$ = 115 μm, $t_s$ = 0.72, and $t_b$ = 0.99. (b) Zoom-in view of (a) in the wavelength range of 1549 nm –1550.7 nm.

## IV. RESONANCE MODE SPLITTING

In this section, we tailor the mode interference in the zig-zag 3WC-SLR resonator to achieve resonance mode splitting with high ERs and low FSRs. The resonance mode splitting with multiple densely spaced resonances can break the intrinsic dependence between the Q factor, FSR, and physical cavity length, thus allowing low FSRs and high Q factors in resonators with a compact footprint. Figure 4(a) shows the power transmission spectrum from Port 2 to Port 4 of the zig-zag 3WC-SLR resonator. The structural parameters are $L_{SLR}$ = $L_{1,2,3,4}$ = 115 μm, $t_s$ = 0.72, and $t_b$ = 0.99, which are designed in order to achieve a CS of about 100 GHz between adjacent split resonances. In Fig. 4(a), CS1 = 98.33 GHz and CS2 = 102.26 GHz. There are two split resonances within a FSR of ~ 200.59 GHz. Figure 4(b) shows a zoom-in view of Fig. 4(a) in the wavelength range of 1549 nm – 1550.7 nm. The IL, Q factor, ER1, and ER2 of the two split resonances in Fig. 4(b) are ~2.02 dB, ~6.03 × $10^4$, ~24.65 dB, and ~27.55 dB, respectively.

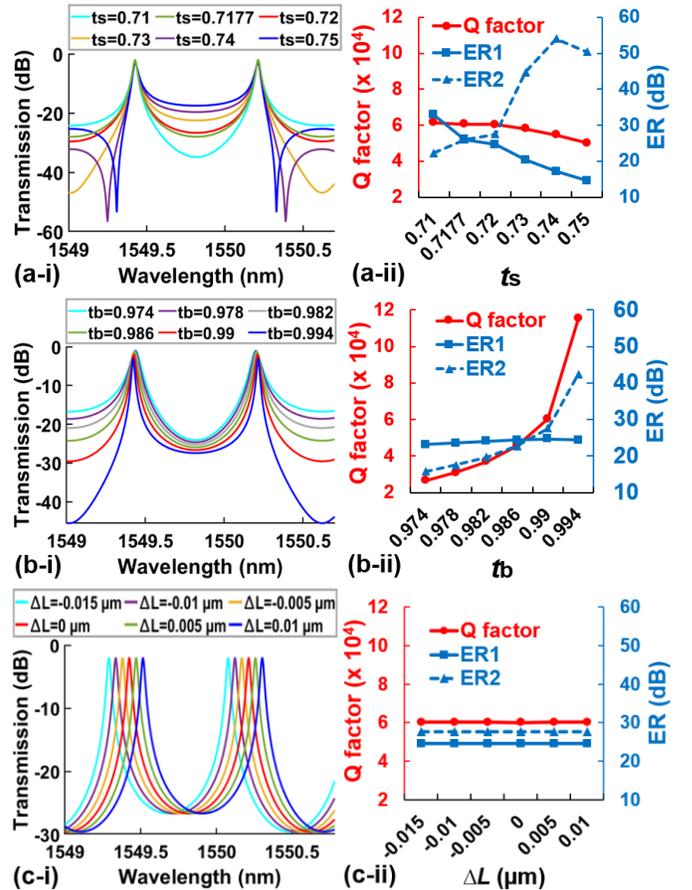

Fig. 5. (a-i) Power transmission spectra of the zig-zag 3WC-SLR resonator for various $t_s$ for input from Port 2 to Port 4 when $t_b$ = 0.99 and $L_{SLR} = L_{1,2,3,4}$ = 115 μm. (a-ii) Calculated Q factor and ERs (ER1 and ER2) as functions of $t_s$ for the transmission spectra in (a-i). (b-i) Power transmission spectra and (b-ii) the corresponding IL and ERs for different $t_b$ when $t_s$ = 0.72 and $L_{SLR}$ = $L_{1,2,3,4}$ = 115 μm, respectively. (c-i) Power transmission spectra and (c-ii) the corresponding Q factor and ERs for different $L$ when $t_s$ = 0.72, $t_b$ = 0.99, and $L_{SLR}$ = 115 μm, respectively.

We further investigate the impact of varying $t_s$, $t_b$, and $L = L_{1,2,3,4}$ on the Q factor, ER1, and ER2 of the split resonances, all of which are important parameters reflecting the degree of mode splitting. In Figs. 5(a) – (c), we only changed one structural parameter in each figure, keeping the others the same as those in Fig. 4. Figure 5(a-i) shows the spectral response for various $t_s$. The Q factor and ERs (ER1 and ER2) as functions of $t_s$ are depicted in Fig. 5(a-ii). As $t_s$ increases, the Q factor slightly decreases while the ER1 and ER2 change more dramatically, resulting in a change in the spectral response towards that of the Fano resonances in Fig. 2(a). The non-monotonic change in ER2 with $t_s$ follows the trend of the SR in Fig. 3(a-ii) for similar reasons. In particular, ER1 equals to ER2 when $t_s$ = 0.7177. Under this condition, the Q factor and effective FSR are ~6.06 × 10$^4$ and ~100.30 GHz (i.e., half of the FSR in Fig. 4(a)), respectively. To achieve the same FSR, the circumference of a comparable MRR (with the same waveguide geometry and loss) is 690 μm, which is 6 times the length of the SLRs. This highlights the reduced cavity length enabled by the mode splitting in the 3WC-SLR resonator. On the other hand, the Q factor of a comparable MRR with the same FSR and ER is ~6.08 × 10$^4$ – almost the same as that of the zig-zag 3WC-SLR resonator. This indicates that the reduced cavity length did not come at the expense of a significant decrease in Q factor. The spectral response for different $t_b$ are shown in Fig. 5(b-i). The corresponding Q factor and ERs are depicted in Fig. 5(b-ii). The ER1 remains almost unchanged while both the ER2 and the Q factor increase with $t_b$, at the expense of a slightly increased IL. The corresponding results for different Δ$L$ are shown in Fig. 5(c-i) and (c-ii). Following the trend in Fig. 2(c), the filter shape remains unchanged while the resonance redshifts as Δ$L$ increases.

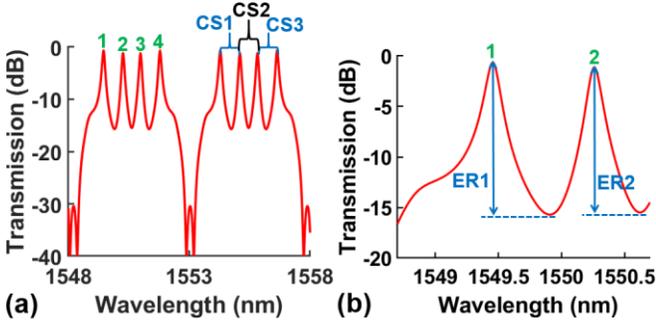

Fig. 6. (a) Power transmission spectrum of the zig-zag 3WC-SLR resonator from Port 1 to Port 3 when $L_{SLR}$ = 115 μm, $L_{1,3}$ = 115 μm, $L_{2,4}$ = 230 μm, and $t_s = t_b = 0.88$. (b) Zoom-in view of (a) in the wavelength range of 1548.7 nm – 1550.7 nm.

The number of split resonances can be changed by varying the length of the connecting bus waveguides. Figure 6(a-i) shows the power transmission spectrum from Port 1 to Port 3 of the zig-zag 3WC-SLR resonator. The structural parameters are $L_{SLR}$ = 115 μm, $L_{1,3}$ = 115 μm, $L_{2,4}$ = 230 μm, and $t_s = t_b$ = 0.88. Clearly, there are four split resonances in each FSR. The CSs between the split resonances are CS1 = CS3 = 100.46 GHz and CS2 = 90.37 GHz. Figure 6(b) shows a zoom-in view of Fig. 6(a) in the wavelength range of 1548.7 nm – 1550.7 nm. To quantitatively analyze the improvement in the performance of the multiple split resonances, in Table III we further compare the resonance mode splitting in the zig-zag 3WC-SLR resonator with that of five cascaded SLRs [163] which can also generate four split resonances. As compared with the five cascaded SLRs that only include standing-wave (SW) filter elements, the mode interference between the SW and travelling-wave (TW) filter elements in the zig-zag 3WC-SLR resonator yields higher ERs for the split resonances, a smaller difference between the ERs of the split resonances, and fewer required SLRs.

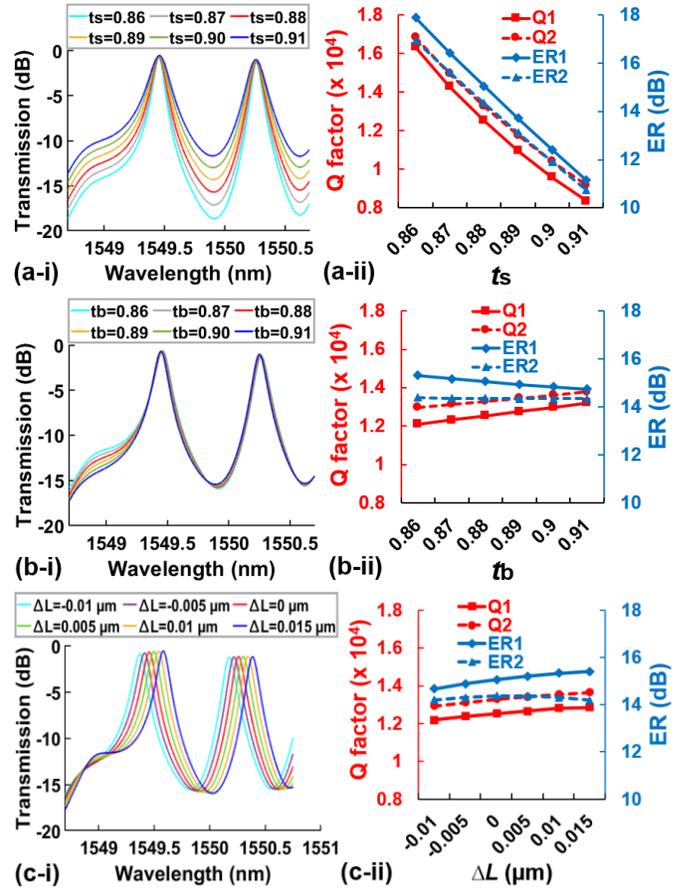

Fig. 7. (a-i) Power transmission spectra of the zig-zag 3WC-SLR resonator for various $t_s$ for input from Port 1 to Port 3 when $t_b$ = 0.88, $L_{SLR}$ = 115 μm, $L_{1,3}$ = 115 μm, and $L_{2,4}$ = 230 μm. (a-ii) Calculated Q factors (Q1 and Q2) and ERs (ER1 and ER2) as functions of $t_s$ for the transmission spectra in (a-i). (b-i) Power transmission spectra and (b-ii) the corresponding Q factors and ERs for different $t_b$ when $t_s$ = 0.88, $L_{SLR}$ = 115 μm, $L_{1,3}$ = 115 μm, and $L_{2,4}$ = 230 μm. (c-i) Power transmission spectra and (c-ii) the corresponding Q factors and ERs versus connecting bus waveguides length variation Δ$L$ when $L_{SLR}$ = 115 μm and $t_s = t_b$ = 0.88, respectively.

TABLE III
PERFORMANCE COMPARISON OF ZIG-ZAG 3WC-SLR AND FIVE CASCADED SLR RESONATORS IN ACHIEVING MODE SPLITTING

| | | Five cascaded SLR resonator [a] | Zig-zag 3WC-SLR resonator |
|---|---|---|---|
| Number of split resonances | | 4 | 4 |
| Number of SLRs | | 5 | 3 |
| IL (dB) | IL1, IL4 [b] | 1.88 | 0.65 |
| | IL2, IL3 | 0.91 | 1.13 |
| ER (dB) | ER1, ER4 | 5.03 | 15.05 |
| | ER2, ER3 | 4.55 | 14.35 |
| CS (GHz) | CS1 | 13.8 | 100.46 |
| | CS2 | 16.24 | 90.37 |
| | CS3 | 13.8 | 100.46 |
| Ref. | | [163] | This work |

[a] The structural parameters of the five cascaded SLR resonator are $L_{SLRi}$ ($i$ = 1–5) = 115 μm, $L_i$ ($i$ = 1–4) = 115 μm, and $t_{si}$ ($i$ = 1–5) = 0.88.
[b] ILs of the four split resonances are labelled as IL1–IL4 from left to right. Due to symmetry of the split resonances, there are IL1 = IL4, IL2 = IL3, ER1=ER4, and ER2=ER3.

TABLE IV
STRUCTURAL PARAMETERS OF CLASSICAL FILTERS BASED ON PARALLEL 3WC-SLR RESONATORS

|  | **Butterworth** | **Bessel** | **Chebyshev Type I** | **Chebyshev Type II** | **Elliptic** |
|---|---|---|---|---|---|
| $t_s$ | 0.89 | 0.83 | 0.89 | 0.85 | 0.96 |
| $t_b$ | 0.94 | 0.94 | 0.97 | 0.94 | 0.795 |
| $L_{SLR}$ (µm) | 173 | 173 | 173 | 173 | 692 |
| $L_i$ (µm) | 173 ($i$ = 1, 3) 346 ($i$ = 2, 4) | 173 ($i$ = 1, 3) 346 ($i$ = 2, 4) | 173 ($i$ = 1, 3) 346 ($i$ = 2, 4) | 173 ($i$ = 1, 3) 346 ($i$ = 2, 4) | 346 ($i$ = 1-4) |
| Input / output ports | Port 2 / Port 3 | Port 2 / Port 3 | Port 2 / Port 3 | Port 2 / Port 4 | Port 2 / Port 4 |
| FSR (GHz) [a] | 100.004 | 100.004 | 100.004 | 100.004 | 100.004 |

[a] The structural parameters are designed in order to achieve a FSR of ~100.004 GHz in the C band to meet the ITU-T spectral grid standard G694.1 [170].

In Figs. 7(a) − (c), we investigate the impact of $t_s$, $t_b$, and $\Delta L$ on the performance of the resonance mode splitting based on the zig-zag 3WC-SLR resonator in Fig. 6. We only changed one structural parameter in each figure, keeping the others the same as those in Fig. 6 (a). The power transmission spectra for different $t_s$ and $t_b$ are shown in Figs. 7(a-i) and (b-i), respectively. The corresponding Q factors (Q1 and Q2) and ERs (ER1 and ER2) for the first two resonances from the left side are shown in Figs. 7(a-ii) and (b-ii) respectively. In Fig. 7(a), all the Q factors and ERs decrease with $t_s$, along with slightly decreased ILs. In Fig. 7(b), as $t_b$ increases, the difference between the two Q factors as well as the difference between the two ERs gradually decrease, resulting in a more symmetric resonance line-shape, which is desirable for reducing filtering distortions. Figures 7(c-i) and (c-ii) compares the corresponding results for various $\Delta L$. As $\Delta L$ increases, the ER1 slightly increase and ER2 slightly decrease while both Q factors slightly increase, which make the filter shapes more asymmetric.

## V. HIGH PERFORMANCE CLASSICAL FILTERS

In this section, we tailor the mode interference in the parallel 3WC-SLR resonator to realize classical filters including Butterworth, Bessel, Chebyshev, and elliptic filters, that all exhibit broad filtering bandwidths and high ERs. The spectral response of these practical filters (solid lines) together with the ideal passband filter (dashed line) are shown in Fig. 8. The Butterworth filter has a flat passband response, while the Bessel filter has a linear phase response over the passband. The Chebyshev filters have either passband ripples (Type I) or stopband ripples (Type II) together with a flat response in the opposite band, resulting in a steeper roll-off than the Butterworth filter. The elliptic filter has both passband and stopband ripples that yields the steepest roll-off among the four types of filters [153].

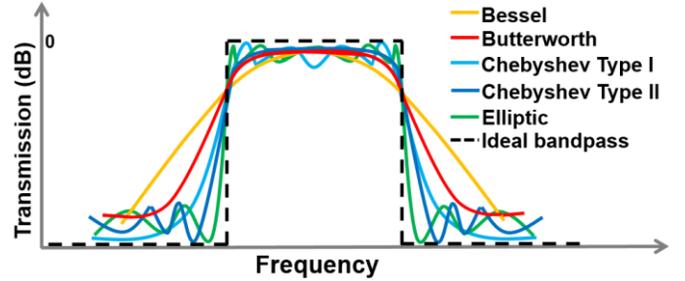

Fig. 8. Spectral responses of classical bandpass filters (solid lines) including Butterworth, Bessel, Chebyshev (Type I and Type II), and elliptic filters. The spectral response of an ideal bandpass filter (dashed lines) is also shown for comparison.

Figure 9(a) shows the power transmission spectrum and corresponding group delay response of the parallel 3WC-SLR resonator from Port 2 to Port 3. As can be seen, there is Butterworth filter shape with flat-top passbands arising from coherent mode interference within the parallel 3WC-SLR resonator, which can be used for low-distortion signal filtering in optical communication systems [171, 172]. The structural parameters are provided in Table IV. The IL, ER, and 3-dB bandwidth (BW) of the Butterworth filter are ~1.71 dB, ~7.12 dB, and ~28.08 GHz, respectively. We then investigate the impact of $t_s$, $t_b$, and $\Delta L$ on the performance of the Butterworth filter. The results are shown in Figs. 9(b) – (d), respectively. In Fig. 9(b), the bandwidth of the passband increases with $t_s$, together with slightly degraded filtering flatness. In Fig. 9(c), the resonance is split, with the spectral range between the split resonances increasing with $t_b$. This indicates that the Butterworth filter shape gradually transitions to a Chebyshev Type I filter shape with improved roll-off and degraded flatness. In Fig. 9(d), as $\Delta L$ increases, the filter shape remains unchanged while the resonance redshifts, showing similar trends to the Fano resonances in Fig. 3(c-i) and the resonance mode splitting in Fig. 5(c-i). This reflects the high fabrication tolerance and the feasibility to realize tunable Butterworth filters.

The spectral and group delay responses of the Bessel filter based on the parallel 3WC-SLR resonator are shown in Fig. 10(a). The input and output ports are Port 2 and Port 3, respectively, the same as those for the Butterworth filter in Fig. 9. The structural parameters are provided in Table IV. As can be seen, the Bessel filter with a flat-top group delay response is achieved, which is useful for applications such as optical buffering and delay lines [173, 174]. The group delay response versus $t_s$, $t_b$, and $\Delta L$ are shown in Figs. 10(b) – (d), respectively. In Fig. 10(b), the bandwidth of the group delay response increases with $t_s$, at the expense of a decreased maximum group delay value and degraded flatness. In Fig. 10(c), the maximum group delay on both sides increases with $t_b$, while the group delay at the center wavelength shows the opposite trend, resulting in higher unevenness in the group delay response. In Fig. 10(d), the group delay response remains unchanged but redshifts as $\Delta L$ increases.

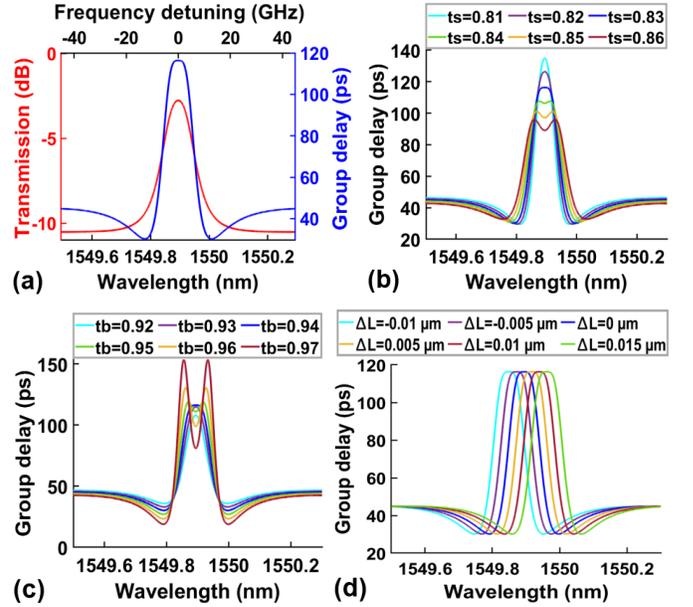

Fig. 10. Bessel filter based on the parallel 3WC-SLR resonator. (a) Power transmission spectra and group delay response from Port 2 to Port 3 when the structural parameters are $L_{SLR}$ = 173 µm, $L_{1,3}$ = 173 µm, $L_{2,4}$ = 346 µm, $t_s$ = 0.83, and $t_b$ = 0.94. (b) – (d) Group delay responses versus $t_s$, $t_b$, and $\Delta L$, respectively. The structural parameters are kept the same as those in (a) except for the varied parameters.

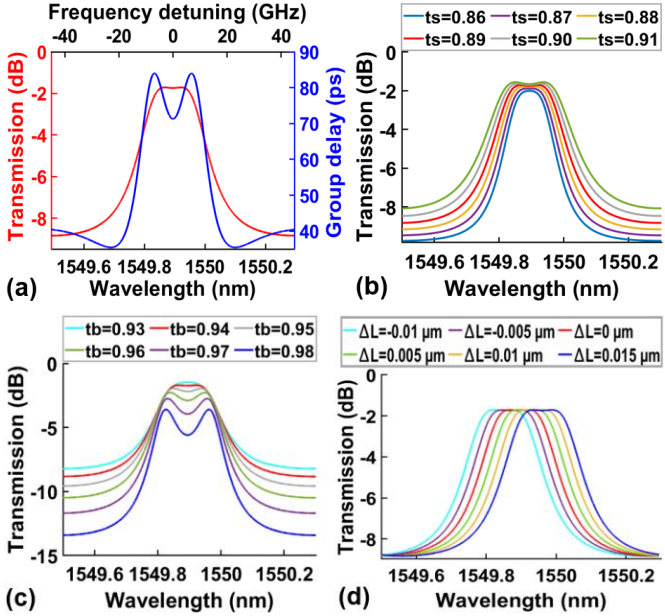

Fig. 9. Butterworth filter based on the parallel 3WC-SLR resonator. (a) Power transmission spectra and group delay response from Port 2 to Port 3. The structural parameters are $L_{SLR}$ = 173 µm, $L_{1,3}$ = 173 µm, $L_{2,4}$ = 346 µm, $t_s$ = 0.89, and $t_b$ = 0.94. (b) – (d) Power transmission spectra versus $t_s$, $t_b$, and $\Delta L$, respectively. The structural parameters are kept the same as those in (a) except for the varied parameters.

Figure 11(a) shows the power transmission spectrum and group delay response of the Chebyshev Type II filter based on the parallel 3WC-SLR resonator. The input port is Port 2, which is the same as those for the Butterworth and Bessel filters, while the output port is changed from Port 3 to Port 4. The structural parameters are provided in Table IV. Clearly, there is a Chebyshev Type II filter shape with equal stopband ripples and a flat response in the passband. The IL, maximum stopband ripple, ER, and 3-dB BW are ~1.26 dB, ~0.037 dB, 27.33 dB, and ~18.77 GHz, respectively. This filter function with a very flat passband and strongly rejected band can be useful for cleaning and extracting channels from crosstalk in a WDM optical communications system, respectively [175]. The power transmission spectra versus $t_s$, $t_b$, and $\Delta L$ are shown in Figs. 11(b) – (d), respectively. As shown in Figs. 11(b) and (c), by increasing $t_s$ and keeping constant $t_b$ or vice versa, the notch depth of the single resonance first increases and then the single resonance is split with an increased spectral range between the split resonances. This is a typical phenomenon for resonance mode splitting, which has also been observed in Refs. [149, 176]. In Fig. 11(d), the filter shape remains unchanged but redshifts as $\Delta L$ increases, following the trends for previous filters.

Finally, we tailor the mode interference in the parallel 3WC-SLR resonator to realize elliptic filters. As compared with the Butterworth and Chebyshev type I filters that have all-pole transfer functions, elliptic filters include both poles and zeros in the transfer function, which can provide higher stopband extinction levels and better roll-off [160]. Figure 12(a) shows the power transmission spectrum and group delay of the parallel 3WC-SLR resonator from Port 2 to Port 4. There is an elliptic filter shape with ripples in both the passband and stopband. The structural parameters are also provided in Table IV. The IL and notch depth of filter are ~0.59 dB and ~31.5 dB, respectively. The maximum passband and stopband ripples are ~1.56 dB and ~18.25 dB, respectively. The ripples in the passband and stopband make the filtering roll-off steeper. These ripples together with the steep roll-off result in a reasonable trade-off between the minimum signal degradation and the maximum noise/interference rejection [177]. The power transmission spectra versus $t_s$, $t_b$, and $\Delta L$ are shown in Figs. 12(b) – (d), respectively. In Figs. 12(b) and (c), the evolution of the split resonance is similar to that observed in Figs. 11 (b) and (c). In Fig. 12(d), unlike the trends for previous filters that exhibit an unchanged filter shape when $\Delta L$ is varied, the filter shape shows a slight asymmetry in the stop-band when $\Delta L$ is away from 0. This is mainly due to the asynchronous feedback in the elliptic filter, which results in asymmetrically located zeros around the center frequency [178].

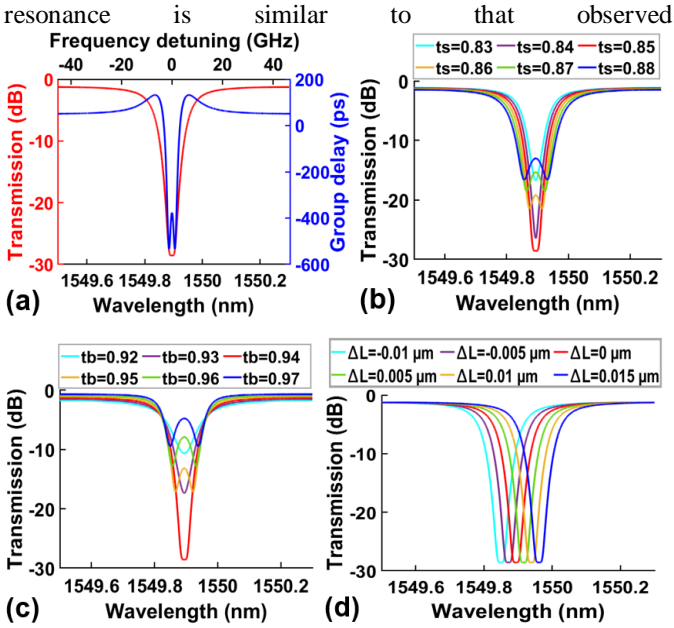

Fig. 11. Chebyshev Type II filter based on the parallel 3WC-SLR resonator. (a) Power transmission spectra and corresponding group delay response from Port 2 to Port 4 when $L_{SLR}$ = 173 μm, $L_{1,3}$ = 173 μm, $L_{2,4}$ = 346 μm, $t_s$ = 0.85, and $t_b$ = 0.94. (b) – (d) Power transmission spectra versus $t_s$, $t_b$, and $\Delta L$, respectively. The structural parameters are kept the same as those in (a) except for the varied parameters.

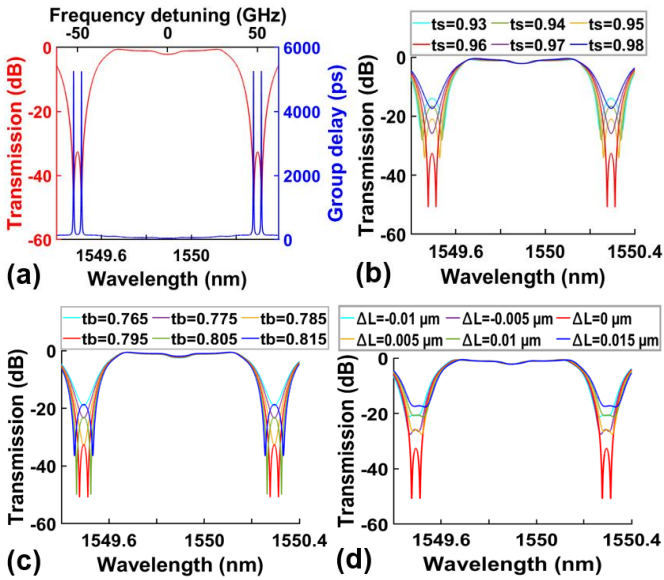

Fig. 12. Elliptic filter based on the parallel 3WC-SLR resonator. (a) Power transmission spectrum and corresponding group delay response from Port 2 to Port 4. The structural parameters are $L_{SLR}$ = 692 μm, $L_{1,2,3,4}$ = 346 μm, $t_s$ = 0.96, and $t_b$ = 0.795. (b) – (d) Power transmission spectra versus $t_s$, $t_b$, and $\Delta L$, respectively. The structural parameters are kept the same as those in (a) except for the varied parameters.

## VI. CONCLUSIONS

We have theoretically investigated advanced multi-functional integrated photonic filters based on 3WC-SLR resonators. Mode interference in the 3WC-SLR resonators is tailored to achieve high performance filtering functions including optical analogues of Fano resonances with ultrahigh ERs and SRs, resonance mode splitting with high ERs and low FSRs, and classical Butterworth, Bessel, Chebyshev, and elliptic filters. The requirements for practical applications are considered in our designs, together with detailed analysis of the impact of structural parameters and fabrication tolerances. This work highlights the 3WC-SLR resonators as a powerful and versatile approach to flexible spectral engineering for a diverse range of applications.

11[102] ...harmonic generation in Si, Ge, and GaAs," *Phys. Rev. B*, vol. 41, no. 3, pp. 1542-1560, 1990.

[103] D. J. Moss, H. M. van Driel, and J. E. Sipe, "Third harmonic generation as a structural diagnostic of ion implanted amorphous and crystalline silicon," *Appl. Phy. Lett.*, vol. 48, no. 17, pp. 1150, 1986.

[104] M.D. Pelusi, F. Luan, E. Magi, M.R.E. Lamont, D. J. Moss, B.J. Eggleton, J.S. Sanghera, L. B. Shaw, and I.D. Aggarwal, "High bit rate all-optical signal processing in a fiber photonic wire", Optics Express vol.16, 11506-11512 (2008).

[105] V. G. Ta'eed, *et al.*, "Self-phase modulation based integrated optical regeneration in chalcogenide waveguides", *IEEE Journal of Selected Topics in Quantum Electronics*, vol. 12, no. 3, pp. 360-370, 2006.

[106] L.Fu, M.Rochette, V. Ta'eed, I.C.M.Littler, D.J. Moss and B.J.Eggleton, "Investigation of self-phase modulation based optical regeneration schemes in single mode As2Se3 chalcogenide glass fiber", Optics Express, vol. 13, 7637 (2005).

[107] V.G. Ta'eed et al., "Integrated all-optical pulse regeneration in chalcogenide waveguides", Optics Letters, vol. 30, 2900 (2005).

[108] V.G. Ta'eed et al., "All Optical Wavelength Conversion via Cross Phase Modulation in Chalcogenide Glass Rib Waveguides", Optics Express, vol. 14, 11242 (2006).

[109] V.G. Ta'eed et al., "Self-phase modulation based integrated optical regeneration in chalcogenide waveguides", IEEE Journal of Selected Topics in Quantum Electronics, Vol.12, 360 (2006).

[110] M. Rochette, L.B. Fu, V.G. Ta'eed, I.C.M. Littler, D.J. Moss, B.J. Eggleton, "2R Optical Regeneration: Beyond Noise Compression to BER Reduction", IEEE Journal of Selected Topics in Quantum Electronics, vol. 12, 736 (2006).

[111] V.G. Ta'eed, L.Fu, M.Rochette, I.C. M. Littler, D.J. Moss, B.J. Eggleton, "Error-Free All-Optical Wavelength Conversion in Highly Nonlinear As$_2$Se$_3$ Chalcogenide Glass Fiber", Optics Express, vol. 14, 10371 (2006).

[112] V.G. Ta'eed, N. Baker, L. Fu, K. Finsterbusch, M.R.E. Lamont, H.C. Nguyen, D.J. Moss, and B.J. Eggleton, Y. Choi, S. Madden, B. Luther-Davies "Ultrafast all-optical chalcogenide glass photonic circuits", Optics Express, vol. 15, 9205 (2007).

[113] C. Monat et al., "Slow light enhanced nonlinear optics in dispersion engineered slow-light silicon photonic crystal waveguides", IEEE Journal of Selected Topics in Quantum Electronics, vol. 16, no. 1, pp. 344-356 (2010).

[114] B. Corcoran, C. Monat, C. Grillet, M.Pelusi, T. P. White, L. O'Faolain, T. F. Krauss, B. J. Eggleton, and D. J. Moss, "Optical signal processing on a silicon chip at 640Gb/s using slow-light", Optics Express, vol. 18, no. 8, 7770-7781 (2010).

[115] V. G. Ta'eed, *et al.*, "Self-phase modulation based integrated optical regeneration in chalcogenide waveguides", *IEEE Journal of Selected Topics in Quantum Electronics*, vol. 12, no. 3, pp. 360-370, 2006.

[116] M. Shokooh-Saremi, *et al.*, "High performance Bragg gratings in chalcogenide rib waveguides written with a modified Sagnac interferometer: experiment and modeling", *Journal of the Optical Society of America B (JOSA B)*, vol. 23, no. 7, pp. 1323-1331, 2006.

[117] M. R. E. Lamont, *et al.*, "Error-free wavelength conversion via cross phase modulation in 5 cm of As$_2$S$_3$ chalcogenide glass rib waveguide", *Electronics Letters*, vol. 43, pp. 945-947, 2007.

[118] T.Monro, D.J.Moss, M. Bazylenko, C. Martijn de Sterke, and L. Poladian, "Observation of self-trapping of light in a self written channel in photosensitive glass", Physical Review Letters, vol. 80, 4072 (1998).

[119] T.Ido, H.Sano, D.J.Moss, S.Tanaka, and A.Takai, "Strained InGaAs/InAlAs MQW electroabsorption modulators with large bandwidth and low driving voltage", IEEE Photonics Technology Letters, Vol. 6, 1207 (1994).

[120] E Ghahramani, DJ Moss, JE Sipe, "Linear and nonlinear optical properties of (GaAs) m/(AlAs) n superlattices", Physical Review B, Vol. 43, No. 11, 9269 (1991).

[121] D. J. Moss, et al., "Ultrafast all-optical modulation via two-photon absorption in silicon-insulator waveguides," Electronics Letters, vol. 41, no. 6, pp. 320-321, 2005. DOI:10.1049/el:20058051

[122] M. R. E. Lamont, et al., "Two-photon absorption effects on self-phase-modulation-based 2R optical regeneration," Photonics Technology Letters, vol. 18, no. 10, pp. 1185-1187, 2006. DOI:10.1109/LPT.2006.874718.

[123] A. Tuniz, G. Brawley, D. J. Moss, and B. J. Eggleton, "Two-photon absorption effects on Raman gain in single mode As2Se3 chalcogenide glass fiber," Optics Express, vol. 16, no. 22, pp. 18524-18534, 2008.

[124] J. Wu, L. Jia, Y. Zhang, Y. Qu, B. Jia, and D. J. Moss, "Graphene oxide: versatile films for flat optics to nonlinear photonic chips", Advanced Materials, vo. 33, no. 3, Article:2006415, pp.1-29 (2021). DOI:10.1002/adma.202006415.

[125] L. Jia, J. Wu, Y. Yang, Y. Du, B. Jia, D. J. Moss, "Large Third-Order Optical Kerr Nonlinearity in Nanometer-Thick PdSe$_2$ 2D dichalcogenide Films: Implications for Nonlinear Photonic Devices", ACS Applied Nano Materials, Vol. 3, No. 7, 6876–6883 June 29 (2020). DOI:10.1021/acsanm.0c01239.

[126] L. Jia et al., "BiOBr nanoflakes with strong nonlinear optical properties towards hybrid integrated photonic devices", Applied Physics Letters (APL) Photonics, vol. 4, 090802 (2019).

[127] Y. Zhang et al., "Enhanced Kerr nonlinearity and nonlinear figure of merit in silicon nanowires integrated with 2D graphene oxide films", ACS Applied Materials and Interfaces, vol. 12, no. 29, 33094−33103 June 29 (2020). DOI:10.1021/acsami.0c07852

[128] Y. Qu et al, "Enhanced nonlinear four-wave mixing in silicon nitride waveguides integrated with 2D layered graphene oxide films", Advanced Optical Materials, Vol. 8, No. 21, 2001048 (2020).

[129] J. Wu et al., "Graphene oxide waveguide polarizers and polarization selective micro-ring resonators", Laser and Photonics Reviews, Vol. 13, No. 9, 1900056 (2019).

[130] Y.Yang et al., "Invited Article: Enhanced four-wave mixing in waveguides integrated with graphene oxide",Applied Physics Letters (APL) Photonics, vol. 3, 120803 (2018).

[131] M. F. Limonov *et al.*, "Fano resonances in photonics," *Nat. Photon..*, vol. 11, pp. 543-554, Sep. 2017.

[132] E. Kamenetskii, A. Sadreev, and A. Miroshnichenko, "*Fano resonances in optics and microwaves: physics and applications*," Springer International Publishing, 2018.

[133] E. Miroshnichenko, S. Flach, and Y. S. Kivshar, "Fano resonances in nanoscale structures," *Rev. Mod. Phys.,* vol. 82, no. 3, pp. 2257-2298, Aug. 2010.

[134] C. Grillet *et al.*, "Characterization and modeling of Fano resonances in chalcogenide photonic crystal membranes," *Opt. Express,* vol. 14, no. 1, pp. 369-376, Jan. 2006.

[135] U. Fano, "Sullo spettro di assorbimento dei gas nobili presso il limite dello spettro d'arco," *Il Nuovo Cimento (1924-1942),* vol. 12, no. 3, pp. 154-161, Mar. 1935.

[136] U. Fano, "Effects of configuration interaction on intensities and phase shifts," *Phys. Rev.,* vol. 124, no. 6, pp. 1866-1878, Dec. 1961.

[137] L. Stern, M. Grajower, and U. Levy, "Fano resonances and all-optical switching in a resonantly coupled plasmonic–atomic system," *Nat. Commun.,* vol. 5, no. 1, pp. 1-9, Sep. 2014.

[138] Y. Deng *et al.*, "Tunable and high-sensitivity sensing based on Fano resonance with coupled plasmonic cavities," *Sci. Rep.,* vol. 7, no. 1, pp. 1-7, Sep. 2017.

[139] B. S. Luk'yanchuk, A. E. Miroshnichenko, and Y. S. Kivshar, "Fano resonances and topological optics: an interplay of far- and near-field interference phenomena," *J. Opt.,* vol. 15, no. 7, May 2013, Art. no. 073001.

[140] B. Peng *et al.*, "What is and what is not electromagnetically induced transparency in whispering-gallery microcavities," *Nat. Commun.,* vol. 5, no. 1, pp. 1-9, Oct. 2014.

[141] Q. Li *et al.*, "Coupled mode theory analysis of mode-splitting in coupled cavity system," *Opt. Express,* vol. 18, no. 8, pp. 8367-8382, Apr. 2010.

[142] M. C. Souza *et al.*, "Spectral engineering with coupled microcavities: active control of resonant mode-splitting," *Opt. Lett.,* vol. 40, no. 14, pp. 3332-5, Jul 2015.

[143] L. A. M. Barea *et al.*, "Silicon technology compatible photonic molecules for compact optical signal processing," *Appl. Phys. Lett.,* vol. 103, no. 20, 2013, Art. no. 201102.

[144] L. Zhou, T. Ye, and J. Chen, "Coherent interference induced transparency in self-coupled optical waveguide-based resonators," *Opt. Lett.,* vol. 36, no. 1, pp. 13-15, Jan. 2011.

[145] C. M. Gentry, X. Zeng, and M. A. Popović, "Tunable coupled-mode dispersion compensation and its application to on-chip resonant four-wave mixing," *Opt. Lett.,* vol. 39, no. 19, pp. 5689-5692, Oct. 2014.
11